%Paper: hep-th/9206064
%From: RCM@hep.physics.mcgill.ca
%Date: Tue, 16 Jun 1992 13:41:14 -0400 (EDT)

\input jnl.tex

\def\part{\partial}
\def\D{{\cal D}}

\def\cit#1{[\cite{#1}]}

\def\al{\alpha}
\def\ga{\gamma}

\def\Ga{\Gamma}

\def\be{\beta}
\def\veps{\varepsilon}
\def\kap{\kappa}
\def\ka{\kappa}

\ignoreuncited
\rightline{McGill/92--25}
\rightline{hep-th/9206064}

\font\ti=cmr7 scaled \magstep5
\title{{\ti What unitary matrix models are not?}}
\author{{\rm Ren\'e Lafrance\footnote{${}^1$}{lafrance@physics.mcgill.ca}
and
Robert C. Myers\footnote{${}^2$}{rcm@physics.mcgill.ca}}}
\affil{\rm Physics Department, McGill University
Ernest Rutherford Building
Montr\'eal, Qu\'ebec H3A 2T8
Canada}
\abstract{We report results of two investigations of the
double-scaling equations for
the unitary matrix models. First, the fixed area partition functions have
all positive coefficients only for the first four critical points.
This implies that the critical points at $k\ge5$ describe non-unitary
continuum theories.
Secondly, we examine a conjectured connection to branched polymers,
but find that the scaling solutions of the unitary models do not agree
with those of a particular model describing branched polymers.}

\endtitlepage
\baselineskip=14truept
\centerline{1. Introduction}
\medskip
Recently, there has been a great deal of interest in random matrix models as a
lattice regulator of two-dimensional Euclidean gravity. The primary stimulus
for this activity was the discovery of the double-scaling limit\cit{double},
which yields results to all orders in the genus expansion.
Following the original work\cit{double}, which was based on hermitian
matrix models, the techniques were extended to find double-scaling solutions
for a host of models based on different matrix ensembles. Some of these
models\cit{25} simply reproduce two-dimensional gravity, and the
Korteweg-deVries hierarchy, as in the
hermitian case. Other models correspond to new physical systems,
including: unoriented surfaces\cit{26}, open strings\cit{27}, strings in
one-dimension\cit{28}, branched polymers\cit{arley,31}, and gravity coupled
to new world-sheet matter systems\cit{29,isin}. Double-scaling solutions
were also found for models where the physical interpretation is
unknown\cit{unit,30,bou}. In particular, Ref. \cit{unit} considered an
integral over unitary matrices, and found double-scaling
solutions where the KdV hierarchy is replaced by the
modified-KdV hierarchy.\footnote
{$^\dagger$}{Ref. \cit{cdm} has shown that these equations are embedded
in an extended system of Zakharov-Shabat flows. We restrict our attention
to the original string equations of Ref. \cit{unit}.} Originally, it was
conjectured that the corresponding continuum theory would
be two-dimensional supergravity.\footnote{$^\ddagger$}{This conjecture
was made independently by a number of individuals including V. Periwal,
E. Witten and N. Seiberg -- see also Ref. \cit{cdm}.}
Ref.~\cit{cdm} also suggests
that the correct identification may be to a dense
phase of branched polymers, or
a topological field theory. The models solved in \cit{30} appear to give
analogous theories, which also include contributions from unoriented
surfaces, and those in \cit{bou} provide comparable open string theories.

Correctly identifying the continuum theory described by a certain
double-scaling solution is a challenging task. It is complicated by
the fact that many different matrix models will produce the same
scaling solutions (see for example \cit{25}). In fact, one can arrange
to produce mKdV critical points in the Hermitian matrix model\cit{cut2},
as well as the KdV critical points in the unitary matrix model\cit{tan}.
At present, it is expected that double-scaling solutions in the KdV
series correspond to pure gravity at $k$=2, and gravity coupled to
nonunitary conformal theories with central charge $c=1- {3(2k-3)\over2k-1}$
for $k\ge3$. This identification is only completely clear for $k=2$, and
for
$k=3$, where a mapping of the lattice theory to the Lee-Yang edge
singularity has been made\cit{staud,isin}. An indication that these models
correspond to non-unitary continuum theories is the fact that in
the genus expansion of the fixed area partition functions for
$k\ge3$, the coefficients become negative for some large genus\cit{big}.
Finding similar results for the fixed area partition functions
for the mKdV series, will provide a significant clue in identifying
the correct continuum theories. The positivity of all the coefficients has been
established analytically for $k=1$, and numerical investigations suggest
that positivity also holds for $k=2$ and 3\cit{unit}.
In Section 2, we examine this question for general critical points
in the mKdV series, using the techniques presented in Ref. \cit{paul}.
We find that positivity of the coefficients only holds for
$k=1,2,3,4.$

Section 3 addresses the possible connection of
double-scaling solutions in the mKdV series to branched polymers\cit{cdm}.
Further support for this conjectured connection may be drawn from the
result that in the analysis of \cit{trip}, it was found that the
scaling function for a particular model of branched
polymers\cit{arley,31} satisfies the Painlev\'e II
equation, the first
equation in the mKdV series. We examine whether or not this connection
survives at the higher order critical points of the branched polymer
model, by directly comparing the scaling functions. Our conclusion
is that no such relation exists at the higher critical points. Thus
the mKdV series is not related to the branched polymer theories described
by these particular models\cit{arley,31}.

\bigskip

\centerline{2. Positivity}
\smallskip

The double-scaling solutions of the unitary matrix model found by
Shevitz and Periwal\cit{unit} are characterized by a free energy $F$,
given by $\part_x^2 F=f^2$. The scaling function $f$ at the $k$'th critical
point satisfies a string equation of the form
$$ x=M_k\ \D^k\!\cdot I\eqno(string)$$
where the recursion operator may be written as
$$\D=-{1\over4f}\part^2_x f +\part^{-1}_x f\,\part_x f\eqno(recur)$$
and the normalization constant is given by $M_k=2^{2k}(k!)^2/(2k)!\ .$
At $k=1$, one finds the Painlev\'e II equation: $f^3-{1\over2}\part_x^2f-xf=0$.
$\D$ given in \(recur) is related to the more familiar recursion operator for
conserved densities of the modified Korteweg-deVries equation by
$\D ={3\over2f} \part_x^{-1} \widehat\D\part_x f$ where
$\widehat\D=-{1\over6}\part_x^2+{2\over3}\part_x f\part^{-1}_x f$
\cit{unit}. $\D$ will prove to be a simpler form for our purposes
below. With the normalization chosen above, the asymptotic expansion
of $f$ for $x\rightarrow\infty$ takes the form
$$ f= x^{1\over2k}\left(1-\sum_{l=1}^\infty f_l (x^{-2-{1\over k}})^l\right)
\ ,\eqno(solu)$$
and similarly for the specific heat, $\part_x^2 F$, one has
$$ F''= x^{1\over k}\left(1-\sum_{l=1}^\infty g_l
(x^{-2-{1\over k}})^l\right) \ .\eqno(ass)$$
Hence one appears to have a genus expansion where the string coupling given
by $\ka^2=x^{-2-{1\over k}}$.

In analogy with Ref. \cit{big}, the fixed area partition functions
(with one point fixed) are defined as the inverse Laplace transform
of $\part_x F$:
$$\rho(A) =\int {dx\over2\pi i}\ \part_x F\ e^{Ax}$$
where the contour runs parallel to the imaginary axis to the right of
any poles. Given the asymptotic expansion \(ass) for the free energy,
one finds\cit{big}
$$\rho(A) = - {A^{-2-{1\over k}}\over\Ga(-{1\over k})}
+\sum_{l=1}^\infty g_l {A^{(l-1)(2+{1\over k})}\over\Ga(
(2+{1\over k})l-{1\over k})}\ \ .$$
Thus the positivity of the coefficients of the fixed area partition
functions requires the positivity of $g_l$, the coefficients appearing
in the expansion of the specific heat.

Ginsparg and Zinn-Justin studied the large order behavior of the expansion
coefficients in various matrix models\cit{paul}.
They showed that the large order behavior is governed by the exponentially
small instanton solutions of the linearized string equations. These
instantons take the form $\veps\approx \kap^\nu e^{-\al/\kap}$.
They are associated to
cuts in the Borel transform of the scaling functions.
The cut ending
closest to the origin governs the large order behavior, and
this in turn corresponds to the instanton(s) with the smallest
value of $|\al|^2$.
We define the leading order behavior of the specific heat given by
\(ass) as $F_o''(\kap)\approx \kap^{-1/(2k+1)}$, and the leading
instanton contribution to the specific heat as $F_i''(\kap)\approx
2f\veps\approx 2\kap^{\nu-1/(4k+2)}e^{-\al/\kap}$. The result of \cit{paul}
is that for ${l\rightarrow\infty}$
$$g_l\propto \int_0^\infty {d\kap\over\kap^{2l+1}}
{F_i''\over F_o''}\propto \int_0^\infty{d\kap\over\kap^{2l+1}} \kap^\ga
e^{-\al/\kap}=\Ga(2l-\ga) \al^{-2l+\ga}\ .$$
For the present analysis, what is relevant is that if $\al$ is real, this
result shows that $g_{l+1}/g_l$ is positive so that $g_l$ remains of a
single sign. On the other hand if $\al$ is complex, the large order
behavior is a sum of contributions from $e^{-\al/\kap}$ and $e^{-\al^*/\kap}$.
In this case, $g_{l+1}/g_l$, and hence $g_l$, has no definite sign
at large orders. Therefore positivity of $g_l$ (and hence
of the coefficients of $\rho(A)$) requires that the leading instantons
are real.

Now consider instanton solutions of \(string) by considering an
infinitesimal perturbation of the asymptotic solution \(solu):
$f=f_o+\veps$ where $f_o\approx x^{1/(2k)}$ and $\veps \propto
e^{-\al x^\be}$ with $Re(\al)>0$. We drop prefactors in $\veps$, which
are powers of $x$, since only $\al$ is relevant for our analysis.
We may also assume $\be>1$, and so to leading order
$\part_x$ and $\part_x^{-1}$ act only on $\veps$. The linearized
string equations reduce to
$$0=\sum_{l=0}^{k-1} \left(-{1\over4} \al^2 \be^2 x^{2\be-2-{1\over k}}
+1\right)^{k-l}\ {x^{-1/2k}\over M_l}\ \veps\ \ .$$
Thus one finds $\be=1+{1\over2k}$, and so the instanton solutions
take the form $\veps\propto e^{-\al/\kap}$, as expected. Given $\be$, the
remaining equation for $\al$ takes the form
$$0=\left(1-{\al^2\be^2\over4}\right)\sum_{l=0}^{k-1}\left(
1-{\al^2\be^2\over4}\right)^{k-1-l} {1\over M_l}\ \ .$$
There is always one real solution given by $\al=2/\be$. To examine the
remaining roots, it is simplest to change variables to $z
=\left(1-{\al^2\be^2\over4}\right)^{-1}$, which must solve
$$0=\sum_{l=0}^{k-1}{z^l\over M_l}\ \ .\eqno(root)$$
If we have another real root, it will govern the large order
behavior if $0<\al^2\be^2<4$. This requires that $0<z<1$, but such roots
clearly can not occur since all the coefficients in \(root) are
positive. Thus $|\al|>2/\be$ for any real solutions, so $\al=2/\be$
for the relevant real instanton. To ensure positivity, one
must ensure that $|\al|>2/\be$ for any complex roots as well. This
corresponds to $Re(z)<1/2$. To proceed further, we resorted to
numerical investigations of \(root). One finds the desired condition
on the complex roots holds for $k=1,2,3,4$.
At $k=5$, one finds the leading roots are the complex conjugate pair
$$z_\pm\approx .54884\pm i\, 1.22727\ .$$
One can
explicitly check that the coefficient at genus 23 becomes negative
for this critical point.
It appears that roots with $Re(z)>1/2$ occur for all $k\ge5$. We have
explicitly checked this result up to $k=200$. For increasing
values of $k$, the leading instantons appear
to be given by complex roots which approach an accumulation point at $z=1$,
such that $|\al^2\be^2|=O(1/k)$.

Therefore for the $k\ge5$ critical points, the leading instantons
have complex $\alpha$ rather than the real solution, $\al=2/\be$,
as was originally conjectured in Ref.~\cit{unit}.
Given the above discussion then, one does not have positivity of the
coefficients in the fixed area partition function for $k\ge5$.
Thus we have the curious result that only
the first four critical points in the double-scaling limit can
correspond to unitary continuum theories. It is tempting to compare
this behaviour to that of the $q$-state Potts model coupled to quantum
gravity. For $q\le4$, the lattice model has a well-defined continuum
limit, while for $q\ge5$, it appears that the continuum limit does
not exist\cit{ball}. Unfortunately, it is clear that there is no
direct connection between these two sets of
systems,\footnote{$^\dagger$}{Clearly, a model with a nonunitary continuum
limit is distinct from one where the continuum limit simply does not exist.
Further the string critical index is $\ga_o=-{1\over2},-{1\over3},
-{1\over5},0$ for the $q=1,2,3,4$ state Potts models,
while
$\ga_o=-1,-{1\over2},-{1\over3},-{1\over4}$
for the first four critical points of the unitary matrix model.}
but the behavior of the unitary matrix models may seem slightly
less remarkable in light of the latter results.

\medskip
\centerline{3. Vector Equations}
\smallskip

Ref.~\cit{cdm} suggested that there may be a connection
between the unitary matrix model results, and a dense phase of branched
polymers. Random vector models have been shown to describe such
a phase of branched polymers\cit{arley,31}. At the $k$'th
critical point in these models, one finds that the string critical
index is $\ga_o=-2/k$, and a spectrum of operators with local
dimensions $d_l=l/k$, where $k=1,2,\ldots$ and $l=0,1,2,\ldots$\ .
Similarly for
the unitary matrix model at the $k$'th critical point, one has
$\ga_o=-1/k$, and $d_l=l/k$. Thus there may be a correspondence between
the mKdV scaling
solutions, and the vector critical points with even $k$,
and their operators
with even $l$. Further support for this suggestion comes from the fact
that in the analysis of \cit{trip}, the Painlev\'e II equation
(with a constant) arises as the string equation for
the $k=2$ vector solution. Painlev\'e II is the first
mKdV string equation. Therefore the scaling functions
at the higher vector critical points (with even $k$)
would be expected to satisfy higher
equations in the mKdV hierarchy. This hypothesis is tested below
by comparing the scaling functions of the $k=4$ vector model with
the $k=2$ unitary matrix model.

Ref.~\cit{arley} introduced $M\times N$ matrix models with a double-scaling
limit, which holds $M$ fixed while $N\rightarrow\infty$. These models can
be regarded as a set of $M$ random vectors, which only interact through
measure terms which are suppresed by a factor of $1/N$\cit{arley}. The
analysis of \cit{trip} introduces $2M$ functions, $g_m$ and $f_m$.
The free energy only depends on the latter through
$F'=-\sum_{m=1}^M f_m$. There are a series of coupled ordinary
differential equations, which these functions must satisfy.
At the $k=2$ point, one can eliminate $g_m$ to leave
$$f_m^3-{1\over2}\part_t^2 f_m -t f_m={1\over2}-m$$
which is the Painlev\'e II equation with a constant. The $k=4$ vector
equations take a more complicated form
$$\eqalign{ t&-f_m^4+12f_m^2g_m-6g_m^2+6f_m^2\part_tf_m-6g_m\part_tf_m\cr
&-3(\part_tf_m)^2-4f_m\part_t^2f_m+2\part_t^2g_m
+\part_t^3f_m=0\cr
m&+4f_m^3g_m-12f_mg_m^2+6f_m^2\part_tg_m-6g_m\part_tg_m\cr
&+2\part_tf_m\part_tg_m+2g_m\part_t^2f_m+4f_m\part_t^2g_m+\part_t^3
g_m=0\cr}\eqno(couple)$$
In this case, we could not eliminate $g_m$ from \(couple), and so we
compared the asymptotic expansion of $f_m$, and the scaling function
solving the $k=2$ mKdV equation
$$f^5-{5\over3}f{f'}^2-{5\over3}f^2f''+{1\over6}f^{(4)}
-xf=h_0 \eqno(k4)$$
where we have allowed for an arbitrary constant $h_0$, on the right
hand side. From \(couple), one finds asymptotic expansions
$$
f_m=t^{1/4}(a_0 - \sum_{l=1}^\infty a_l\ t^{-5l/4})\ \ \ \ {\rm and}
\ \ \ \ g_m=-t^{1/2}\sum_{l=1}^\infty b_l\ t^{-5l/4} $$
while \(k4) yields
$$ f=x^{1/4}(A_0 - \sum_{l=1}^\infty A_l\ x^{-5l/4})\eqno(panda)$$
where $a_0=1=A_0$, and for $l\ge1$, $a_l$ and $A_l$ are functions of $m$
and $h_0$, respectively. A valid comparison can only involve equating
combinations of the coefficients which are independent of the normalizations
of the scaling functions, $f_m$ and $f$, and the scaling variables, $t$
and $x$. For example, one may equate $A_0 A_2/A_1^{\ 2}=a_0 a_2/a_1^{\ 2}$
to find $h_0^{\ 2}=-1-{1\over4m(m-1)}$. Clearly, this is an unsatisfactory
result since it does not yield a finite real result for $h_0$ when $m$ is
a positive integer. Further one finds that it is inconsistent with
equalities involving higher coefficients, $A_1 A_3/A_2^{\ 2}
=a_1 a_3/a_2^{\ 2}$ and $A_2 A_4/A_3^{\ 2}=a_2 a_4/a_3^{\ 2}$.
Thus $f_m$ simply do not satisfy an equation of the form \(k4).

One might think of generalizing \(k4) by also allowing for an extra
perturbation by the Painlev\'e II operator on the right hand side,
$h_1(f^3-{1\over2}f'')$. If $h_1$ is a constant though, this perturbation
changes the character of the asymptotic expansion \(panda) by introducing
new powers: $x^{-1/4},$ $x^{-3/4},\ldots$\ . This problem can be corrected
by multiplying the new perturbation by an overall factor of $f^2$ or $x^{1/2}$,
but further investigation again leads to
inconsistent results when one attempts to equate normalization
independent ratios of the coefficients as above.

Thus there appears to be no
relation between the scaling solutions of the unitary matrix models and
these random vector models, other than the coincidental equality of the
critical index $\ga_o$, and the dimensions of the scaling operators.
This reinforces the difficulty of identifying the correct physical
interpretation for the scaling solutions of the unitary models.
We should state though that our analysis does not rule out a
correspondence with other independent models of branched polymers, such
as those studied in Ref. \cit{poly}.

\bigskip
R.C.M.~gratefully acknowledges useful discussions with V.~Periwal and
M.~Bowick; as well as the hospitality of both the Aspen Center for
Physics, and the Institute for Theoretical Physics at UCSB, at various
stages of this work. This research was supported by NSERC of Canada, and
Fonds FCAR du Qu\'ebec. After this work was completed, we became aware
of \cit{tim}, which offers a world-sheet interpretation of the scaling
solutions of the unitary matrix models in terms of an open-closed string
theory.

\references

\refis{tim} S. Dalley, C.V. Johnson, T.R. Morris and A. W\"atterstam,
`Unitary Matrix Models and 2D Quantum Gravity,' PUPT--1325, SHEP 91/92--19,
G\"oteborg ITP 92--20

\refis{ball} C.F. Baillie and D.A. Johnston, `A numerical test of KPZ
scaling: Potts models coupled to two-dimensional quantum gravity,'
COLO-HEP-269; `Multiple Potts models coupled to two-dimensional
quantum gravity,' COLO--HEP--276

\refis{arley} A. Anderson, R.C. Myers and V. Periwal, {\sl Phys. Lett.}
{\bf 254B} (1991) 89; {\sl Nucl. Phys.} {\bf B360} (1991) 463

\refis{trip} R.C. Myers and V. Periwal, `From polymers to quantum gravity:
triple-scaling in rectangular matrix models,' McGill/92--01,
IASSNS--HEP--91/93

\refis{double} E. Br\'ezin and V.A. Kazakov, {\sl Phys. Lett.} {\bf B236}
(1990) 144; M.R. Douglas and S.H. Shenker, {\sl Nucl. Phys.} {\bf B335}
(1990) 635; D.J. Gross and A.A. Migdal, {\sl Phys. Rev. Lett.} {\bf 64}
(1990) 127

\refis{25} R.C. Myers and V. Periwal, {\sl Phys. Rev.} {\bf D42} (1990)
3600; T.R. Morris, {\sl Nucl. Phys.} {\bf B356} (1991) 703

\refis{26} E. Br\'ezin and H. Neuberger, {\sl Phys. Rev. Lett.} {\bf 65}
(1990) 2098; {\sl Nucl. Phys.} {\bf B350} (1991) 513; G.R. Harris and E.J.
Martinec, {\sl Phys. Lett.} {\bf 245B} (1990) 384; R.C. Myers and
V. Periwal, {\sl Phys. Rev. Lett.} {\bf 64} (1990) 3111; H. Roussel,
McGill Master's thesis (1992)

\refis{27} I.K. Kostov, {\sl Phys. Lett.} {\bf 238B} (1990) 181;
J.A. Minahan, {\sl Phys. Lett.} {\bf 268B} (1991) 29

\refis{bou} J.A. Minahan, {\sl Phys. Lett.} {\bf 265B} (1991) 382

\refis{28} D.J. Gross and N. Miljkovi\'c, {\sl Phys. Lett.} {\bf 238B}
(1990) 217;
E. Brezin, V.A. Kazakov and Al.B. Zamolodchikov, {\sl Nucl. Phys.}
{\bf B338} (1990) 673; P. Ginsparg and J. Zinn-Justin, {\sl Phys. Lett.}
{\bf 240B} (1990) 333

\refis{isin} D.J. Gross and A. Migdal, {\sl Phys. Rev. Lett.} {\bf 64}
(1990) 717

\refis{29} E. Brezin, M. Douglas, V. Kazakov and S. Shenker, {\sl Phys. Lett.}
{\bf 237B} (1990) 43; C. Crnkovic, P. Ginsparg and G. Moore, {\sl
Phys. Lett.} {\bf 237B} (1990) 190; M. Douglas, {\sl Phys. Lett.} {\bf 238B}
(1990) 176

\refis{unit} V. Periwal and D. Shevitz, {\sl Phys. Rev. Lett.} {\bf 64}
(1990) 1326; {\sl Nucl. Phys.} {\bf B344} (1990) 731

\refis{30} R.C. Myers and V.
Periwal, {\sl Phys. Rev. Lett.} {\bf 65} (1990) 1088; M.M. Robinson,
{\sl Phys. Rev.} {\bf D45} (1992) 2872

\refis{31} S. Nishigaki and T. Yoneya, {\sl Nucl. Phys.} {\bf B348} (1991) 787;
P. Di Vecchia, M. Kato and N. Ohta, {\sl Nucl. Phys.} {\bf B357} (1991)
495; R.C. Myers, {\sl Nucl. Phys.} {\bf B368} (1992) 701

\refis{cdm} \u C. Crnkovic, M. Douglas and G. Moore, `Loop Equations and
the Topological Phase of Multi-Cut Matrix Models,' Rutgers preprint
91--36

\refis{cut2} M. Douglas, N. Seiberg and S. Shenker, {\sl Phys. Lett.} {\bf
244B} (1990) 381; \u C. Crnkovic and G. Moore, {\sl Phys. Lett.} {\bf
257B} (1991) 322; P. Mathieu and D. S\'en\'echal, {\sl Phys. Lett.} {\bf
267B} (1991) 475

\refis{tan} K. Demeterfi and C.-I. Tan, {\sl Mod. Phys. Lett.} {\bf
A5} (1990) 1563

\refis{staud} M. Staudacher, {\sl Nucl. Phys.} {\bf B336} (1990) 349

\refis{big} D.J. Gross and A.A. Migdal, {\sl Nucl. Phys.} {\bf B340}
(1990) 333

\refis{paul} P. Ginsparg and J. Zinn-Justin in
{\it Random Surfaces and Quantum Gravity}, O. Alvarez, E. Marinari
and P. Windey, eds., 1990 Carg\`ese workshop; {\sl Phys. Lett.} {\bf
B255} (1991) 189

\refis{poly} B. Duplantier and I.K. Kostov, {\sl Nucl. Phys.} {\bf B340}
(1990) 491; B. Duplantier and H. Saleur, {\sl Nucl. Phys.} {\bf B290}
(1987) 291; H. Saleur, `Polymers and Percolation in Two-Dimensions and
Twisted N=2 Supersymmetry,' Yale preprint YCTP-P38-91

\endreferences
\end